\documentclass[preprint,eqsecnum,aps,nofootinbib]{revtex4}
\usepackage{amsfonts,amsmath,amssymb,amsthm}
\usepackage{latexsym}
\usepackage{bbm,bm}
\usepackage{graphicx}

\newcommand{\ket}[1]{\lvert #1 \rangle}
\newcommand{\bra}[1]{\langle #1 \lvert}
\newcommand{\beq}{\begin{equation}}
\newcommand{\eeq}{\end{equation}}
\newcommand{\beqs}{\begin{eqnarray}}
\newcommand{\eeqs}{\end{eqnarray}}
\begin{document}

\title{Four-Qubit Monogamy and Four-Way Entanglement}

\author{DaeKil Park$^{1,2}$}

\affiliation{$^1$Department of Electronic Engineering, Kyungnam University, Changwon
                 631-701, Korea    \\
             $^2$Department of Physics, Kyungnam University, Changwon
                  631-701, Korea    
                      }

\begin{abstract}
We examine the various properties of the three four-qubit monogamy relations, all of which introduce 
the power factors  in the three-way entanglement to reduce the tripartite contributions. On the analytic ground as much as possible 
we try to find the minimal power factors, which make the monogamy relations hold if the power factors are larger than the minimal powers. 
Motivated to the three-qubit monogamy inequality we also examine whether those four-qubit monogamy relations provide the SLOCC-invariant 
four-way entanglement measures or not. Our analysis indicate that this is impossible provided that the monogamy inequalities are derived 
merely by introducing weighting power factors. 
\end{abstract}

\maketitle

\section{Introduction}
Recently, quantum technology, i.e. technology based on quantum mechanics, attracts much attention to overcome various limitations of classical 
technology such as computational speed of computer and insecurity of cryptography. Quantum entanglement\cite{text,horodecki09} is the most important
physical resource to develop the quantum technology because it plays a crucial role in the various quantum information processing.
In fact, it is used  in quantum teleportation\cite{teleportation}, superdense coding\cite{superdense}, quantum cloning\cite{clon}, and quantum cryptography\cite{cryptography,cryptography2}. It is also quantum entanglement, which makes the quantum computer\footnote{The current status of quantum computer technology was reviewed in Ref.\cite{qcreview}.} outperform the classical one\cite{computer}. Thus, it is very important to understand how to quantify and how to characterize the entanglement.

One of the surprising property of the quantum entanglement arises in its distribution in the multipartite system. It is usually called the 
monogamy property. For example, let us consider the tripartite quantum state $\ket{\psi}_{ABC}$ in the qubit system. 
Authors in  Ref. \cite{ckw} have shown the inequality
\begin{equation}
\label{ckw}
{\cal C}^2_{A|(BC)} \geq {\cal C}^2_{A|B} +  {\cal C}^2_{A|C}
\end{equation}
where ${\cal C}$ is concurrence\cite{woot-98}, one of the entanglement measure for bipartite system.
This inequality, usually called CKW inequality, implies that the entanglement (measured by the squared concurrence) between $A$ and the remaining parties always exceeds 
entanglement between $A$ and $B$ plus entanglement between $A$ and $C$. This means that the more $A$ and $B$ are entangled, the lesser
$A$ and $C$ are entangled. This is why the quantum cryptography is more secure than classical one. The inequality (\ref{ckw}) is strong in a sense that the three-qubit W-state\cite{dur00}
\begin{equation}
\label{w3}
\ket{\mbox{W}_3} = \frac{1}{\sqrt{3}} \left( \ket{001} + \ket{010} + \ket{100} \right)
\end{equation}
saturates the inequality.

Another surprising property of Eq. (\ref{ckw}) is the fact that  the leftover in the inequality 
\begin{equation}
\label{residual}
\tau_{ABC} = {\cal C}^2_{A|(BC)} - \left(  {\cal C}^2_{A|B} +  {\cal C}^2_{A|C} \right),
\end{equation}
quantifies the true three-way entanglement. For general  three-qubit pure state $\ket{\psi} = \sum_{i,j,k=0}^1 \psi_{ijk} \ket{ijk}_{ABC}$ 
the leftover $\tau_{ABC}$, which is called the residual entanglement\footnote{In this paper $\sqrt{\tau_{ABC}}$ is called the three-tangle.}, reduces to
\begin{equation}
\label{tangle3}
\tau_{ABC} = \bigg|2 \epsilon_{i_1 i_2} \epsilon_{i_3 i_4} \epsilon_{j_1 j_2} \epsilon_{j_3 j_4} \epsilon_{k_1 k_3} \epsilon_{k_2 k_4}
          \psi_{i_1 j_1 k_1} \psi_{i_2 j_2 k_2} \psi_{i_3 j_3 k_3} \psi_{i_4 j_4 k_4} \bigg|.
\end{equation}
From this expression one can show that $\tau_{ABC}$ is invariant under  a stochastic local operation and classical 
communication (SLOCC)\cite{bennet00}.

Then, it is natural to ask whether or not such surprising properties are maintained in the  monogamy relation of multipartite system.
Subsequently, the generalization of Eq. (\ref{ckw}) was discussed in Ref. \cite{osborne06-1}.
 As Ref. \cite{osborne06-1} has shown analytically,  the following monogamy relation
\begin{equation}
\label{tofv}
{\cal C}^2_{q_1|(q_2 \cdots q_n)} \geq {\cal C}^2_{q_1|q_2} +  {\cal C}^2_{q_1|q_3} + \cdots +  {\cal C}^2_{q_1|q_n}
\end{equation}
holds in the $n$-qubit pure-state system. However, it is shown that the leftover of Eq. (\ref{tofv}) is not entanglement monotone. 
In order to remove this unsatisfactory feature the authors in Ref. \cite{bai07-1,bai08-1} considered the average leftover of the monogamy relation (\ref{tofv}).
For example, they conjectured that in four-qubit system the following average leftover
\begin{equation}
\label{tangle-1}
\theta_{ABCD} = \frac{\pi_A + \pi_B + \pi_C + \pi_D}{4}
\end{equation}
is a monotone, where $\pi_A = {\cal C}^2_{A|(BCD)} - ({\cal C}^2_{A|B} + {\cal C}^2_{A|C} + {\cal C}^2_{A|D})$ and other ones are derived by 
changing the focusing qubit. Even though $\theta_{ABCD}$ might be an entanglement monotone, it is obvious that it cannot quantify a true four-way 
entanglement because it detects the partial entanglement. For example, $\theta_{ABCD} (g_3) = 3 / 4$, where 
$\ket{g_3} = \ket{\mbox{GHZ}_3} \otimes \ket{0}$ and $\ket{\mbox{GHZ}_3}$ is a three-qubit Greenberger-Horne-Zeilinger (GHZ) state
defined as 
\begin{equation}
\label{ghz3}
\ket{\mbox{GHZ}_3} = \frac{\ket{000} + \ket{111}}{\sqrt{2}}.
\end{equation}

In Ref. \cite{regula14-1} another following multipartite monogamy relation is examined:
\begin{equation}
\label{smonogamy-1}
{\cal C}^2_{q_1|(q_2 \cdots q_n)} \geq \underbrace{\sum_{j=2}^n {\cal C}^2_{q_1|q_j}}_{2-\mbox{partite}} + \underbrace{\sum_{k > j=2}^n \left[ t_{q_1|q_j|q_k} \right]^{\mu_3}}_{3-\mbox{partite}} + \cdots + 
\underbrace{\sum_{\ell = 2}^n \left[ t_{q_1|q_2|\cdots | q_{\ell-1} | q_{\ell+ 1} | \cdots |q_n} \right]^{\mu_{n-1}}}_{(n-1)-\mbox{partite}}.
\end{equation}
In Eq. (\ref{smonogamy-1}) the power factors $\left\{\mu_m \right\}_{m=3}^{n-1}$ are included to regulate the weight assigned to the different
$m$-partite contributions. If all power factors $\mu_m$ go to infinity,   Eq. (\ref{smonogamy-1}) reduces to Eq. (\ref{tofv}). As a tripartite entanglement 
measure the residual entanglement or three-tangle can be used independently. Thus, in four-qubit system Eq. (\ref{smonogamy-1}) reduces to  following 
two different expressions:
\begin{equation}
\label{four-monogamy}
\Delta_j = t_{1|234} - \left( t_{1|2} + t_{1|3} + t_{1|4} \right) - \left( t_{1|2|3}^{(j)} + t_{1|2|4}^{(j)} + t_{1|3|4}^{(j)} \right)   \geq 0 
\hspace{1.0cm} (j = 1, 2, 3)
\end{equation}
where
\begin{eqnarray}
\label{four-monogamy-1}
&& t_{1|234} = {\cal C}^2_{1|234} =  4 \mbox{det} \rho_1    \hspace{4.0cm} t_{i|j} = {\cal C}^2 (\rho_{ij})    \hspace{1.0cm}    \\   \nonumber
&& t_{i|j|k}^{(1)} = \left[ \min_{\{p_n, \ket{\psi_n}\}} \sum_n p_n \sqrt{\tau_{ijk}} (\psi_n) \right]^{\mu_1}     \hspace{1.0cm}
t_{i|j|k}^{(2)} = \left[ \min_{\{p_n, \ket{\psi_n}\}} \sum_n p_n \tau_{ijk} (\psi_n) \right]^{\mu_2}.    
\end{eqnarray}   
In Eq. (\ref{four-monogamy-1}) the tripartite entanglements $t_{i|j|k}^{(j)}$ are expressed explicitly as a convex roof\cite{benn96,uhlmann99-1} for 
mixed states derived by the partial trace of the four-qubit pure states. In particular, the authors of Ref. \cite{regula14-1} conjectured that 
all four-qubit pure states holds $\Delta_1 \geq 0$ when $\mu_1 \geq 3$. Different expression of the monogamy relation was introduced in Ref. \cite{regula16-1},
which is Eq. (\ref{four-monogamy}) with $j=3$, where 
\begin{equation}
\label{additional-1}
 t_{i|j|k}^{(3)} = \left[ \min_{\{p_n, \ket{\psi_n}\}} \sum_n p_n \tau_{ijk}^{1 / q} (\psi_n) \right]^{q}.
 \end{equation}
 The authors of Ref. \cite{regula16-1} conjectured  that $\Delta_3$ with $q = 4$ might be nonnegative for all four-qubit pure states. They also conjectured
 by making use of their extensive numerical tests that all possible second class states\footnote{This is classified as $L_{abc_2}$ in Ref. \cite{fourP-1}.}
\begin{equation}
\label{second-class}
\ket{G} = {\cal N} \left[ \frac{a + b}{2} \left( \ket{0000} + \ket{1111} \right) +  \frac{a - b}{2} \left( \ket{0011} + \ket{1100} \right) + 
c \left( \ket{0101} + \ket{1010} \right) + \ket{0110} \right]
\end{equation}
 and their 
 SLOCC transformation hold $\Delta_3 \geq 0$ when $q \geq 2.42$,
where the parameters $a$, $b$, and $c$ are generally complex, and ${\cal N}$ is a normalization constant given by 
\begin{equation}
\label{normalization1}
{\cal N} = \frac{1}{\sqrt{1 + |a|^2 + |b|^2 + 2 |c|^2}}.
\end{equation} 

The purpose of this paper is two kinds. First one is to find the minimal powers $(\mu_1)_{\min}$, $(\mu_2)_{\min}$, and $(q)_{\min}$ 
which make $\Delta_j \geq 0$ when the corresponding powers are larger than the minimal powers. Second one is to examine whether or not 
the leftovers $\Delta_j \hspace{.3cm} (j = 1, 2, 3)$ can be true four-way SLOCC-invariant entanglement measures like the CKW inequality in three-qubit case. 
In order to explore these issues 
on the analytical ground as much as possible we confine ourselves into the second class state $\ket{G}$. In Sec. II and Sec. III various tangles are 
computed analytically. In fact, the three-tangle of $\ket{G}$ was computed in Ref.\cite{ostr16-1}. Since, however, there is some mistake in Ref.\cite{ostr16-1}, 
we compute $\tau_{ABC}^{1/q} \hspace{.3cm} (q = 1, 2, \cdots)$ of $\ket{G}$ analytically in Sec. III. In Sec. IV we compute $\Delta_j$ analytically 
for few special cases. Exploiting the numerical results we compute the minimal powers for the cases. In Sec. V we compute the minimal powers for more 
general cases. In Sec. VI we examine whether or not $\Delta_j$ with particular powers can be SLOCC-invariant four-way entanglement measures. 
Our analysis indicate that this is impossible provided that the monogamy inequalities are derived 
merely by introducing weighting power factors. In Sec. VII a brief conclusion is given.

\section{one- and two-tangles}

In order to compute the one-tangle we derive the state of the first qubit $\rho_1$:
\begin{equation}
\label{one-tangle-1}
\rho_1 = \mbox{tr}_{234} \ket{G} \bra{G} = \frac{{\cal N}^2}{4 {\cal N}_2^2} \ket{0} \bra{0} +  \frac{{\cal N}^2}{4 {\cal N}_1^2} \ket{1} \bra{1}
\end{equation}
where ${\cal N}_1$ and ${\cal N}_2$ are 
\begin{equation}
\label{normalization2}
{\cal N}_1 = \frac{1}{\sqrt{2 (|a|^2 + |b|^2 + 2 |c|^2)}}         \hspace{1.0cm}
{\cal N}_2 = \frac{1}{\sqrt{2 (2 + |a|^2 + |b|^2 + 2 |c|^2)}}.
\end{equation}
It is easy to show the equality $1 / {\cal N}_1^2 + 1 / {\cal N}_2^2 = 4 / {\cal N}^2$, which guarantees the normalization of $\rho_1$. 
Thus, the one-tangle of $\rho_1$ is given by 
\begin{equation}
\label{one-tangle-2}
t_{1|234} \equiv 4 \mbox{det} \rho_1 = \frac{{\cal N}^4}{4 {\cal N}_1^2 {\cal N}_2^2}.
\end{equation}
In fact, one can show $t_{1|234} = t_{2|134} = t_{3|124} = t_{4|123}$.

In order to compute the two-tangles we derive the two-qubit states, which are obtained by taking the partial trace over the remaining qubits.
The final results can be represented  as the following matrices in the computational basis:
\begin{eqnarray}
\label{two-tangle-1}
&&\rho_{12} = {\cal N}^2 \left(      \begin{array}{cccc}
                                              \frac{|a|^2 + |b|^2}{2}  &  0  &  0  &  \frac{|a|^2 - |b|^2}{2}            \\
                                              0  &  1 + |c|^2  &  c^*  &  0                                                                 \\
                                              0  &  c  &  |c|^2  &  0                                                                        \\
                                               \frac{|a|^2 - |b|^2}{2}   &  0  &  0  &  \frac{|a|^2 + |b|^2}{2}
                                                    \end{array}                                                                                     \right)    \hspace{.5cm}
\rho_{13} = {\cal N}^2    \left(                 \begin{array}{cccc}
                                               \beta  &  0  &  0  &  \alpha                                                              \\
                                               0  &  \gamma + 1  &  \delta^*  &  0                                                   \\
                                               0  &  \delta  &  \gamma  &  0                                                           \\
                                               \alpha  &  0  &  0  &  \beta
                                                               \end{array}                                                                          \right)
                                                                                                                                                                                           \nonumber   \\
&& \hspace{4.0cm}
\rho_{14} = {\cal N}^2    \left(                 \begin{array}{cccc}
                                               \gamma' + 1  &  0  &  0  &  \delta'                                                             \\
                                               0  &  \beta'  &  \alpha' &  0                                                   \\
                                               0  &  \alpha'  &  \beta'  &  0                                                           \\
                                               (\delta')^*  &  0  &  0  &  \gamma'
                                                               \end{array}                                                                          \right)
\end{eqnarray}
where
\begin{eqnarray}
\label{two-tangle-2}
&&\alpha = \mbox{Re} \left[(a+b) c^* \right]  \hspace{.5cm}  \beta = \frac{|a + b|^2}{4} + |c|^2  \hspace{.5cm}
\gamma = \frac{|a-b|^2}{4}   \hspace{.5cm}  \delta = \frac{a - b}{2}                                                                               \\   \nonumber
&&\alpha' = \mbox{Re} \left[(a-b) c^* \right]  \hspace{.5cm}  \beta' = \frac{|a - b|^2}{4} + |c|^2  \hspace{.5cm}
\gamma' = \frac{|a+b|^2}{4}   \hspace{.5cm}  \delta' = \frac{a + b}{2}.
\end{eqnarray}
Following the Wootters procedure\cite{woot-98} one can compute the two-tangles of the two-qubit reduced states
$t_{i|j} = {\cal C}^2 (\rho_{ij})$ straightforwardly. The final expressions of the concurrences can be written as follows:
\begin{eqnarray}
\label{two-tangle-3}
&&{\cal C} (\rho_{12}) = \left\{                      \begin{array}{cc}
                                     {\cal N}^2 \max \left[2 |c| - (|a|^2 + |b|^2 ), 0 \right]  & \hspace{.3cm} |c| \left[\sqrt{1 + |c|^2} + 1 \right] \geq \left\{|a|^2, |b|^2 \right\}                           \\
                                     {\cal N}^2 \max \left[|a|^2 - |b|^2 - 2 |c| \sqrt{1 + |c|^2}, 0 \right]  &  \hspace{.3cm} |a|^2 \geq \left\{|b|^2,  |c| \left[\sqrt{1 + |c|^2} + 1 \right] \right\}     \\
                                     {\cal N}^2 \max \left[|b|^2 - |a|^2 - 2 |c| \sqrt{1 + |c|^2}, 0 \right]  &  \hspace{.3cm} |b|^2 \geq \left\{|a|^2,  |c| \left[\sqrt{1 + |c|^2} + 1 \right] \right\}
                                                           \end{array}                                                                                                                                                                                        \right.                     \nonumber     \\
&&{\cal C}  (\rho_{13}) = \left\{                      \begin{array}{cc}
                                    2 {\cal N}^2 \max \left[ |\delta| - \beta, 0 \right]  & \hspace{.3cm} \sqrt{\gamma (\gamma + 1)} + |\delta| \geq \left\{ \beta + \alpha, \beta - \alpha \right\}                           \\
                                    2 {\cal N}^2 \max \left[ \alpha - \sqrt{\gamma (\gamma + 1)}, 0 \right]  &  \hspace{.3cm} \beta + \alpha \geq \left\{ \beta - \alpha,  \sqrt{\gamma (\gamma + 1)} + |\delta| \right\}      \\
                                    2 {\cal N}^2 \max \left[- \alpha - \sqrt{\gamma (\gamma + 1)}, 0 \right]  &  \hspace{.3cm}  \beta - \alpha \geq \left\{ \beta + \alpha,  \sqrt{\gamma (\gamma + 1)} + |\delta| \right\}
                                                           \end{array}                                                                                                                                                                                          \right.                     \\      \nonumber
&&{\cal C}  (\rho_{14}) = \left\{                      \begin{array}{cc}
                                    2 {\cal N}^2 \max \left[ |\delta'| - \beta', 0 \right]  & \hspace{.3cm} \sqrt{\gamma' (\gamma' + 1)} + |\delta'| \geq \left\{ \beta' + \alpha', \beta' - \alpha' \right\}                           \\
                                    2 {\cal N}^2 \max \left[ \alpha' - \sqrt{\gamma' (\gamma' + 1)}, 0 \right]  &  \hspace{.3cm} \beta' + \alpha' \geq \left\{ \beta' - \alpha',  \sqrt{\gamma' (\gamma' + 1)} + |\delta'| \right\}      \\
                                    2 {\cal N}^2 \max \left[- \alpha' - \sqrt{\gamma' (\gamma' + 1)}, 0 \right]  &  \hspace{.3cm}  \beta' - \alpha' \geq \left\{ \beta' + \alpha',  \sqrt{\gamma' (\gamma' + 1)} + |\delta'| \right\}
                                                           \end{array}                                                                                                                                                                                          \right.                               
\end{eqnarray}
where $a \geq \left\{b, c \right\}$ means $a \geq b$ and $a \geq c$.

\section{three-tangle}

In order to compute the three-tangles we should derive the three-qubit states by taking partial trace over the remaining qubit. 
For example, $\rho_{123}$ can be written as 
\begin{equation}
\label{three-tangle-1}
\rho_{123} \equiv \mbox{tr}_4 \ket{G} \bra{G} = p \ket{\psi_1} \bra{\psi_1} + (1 - p) \ket{\psi_2} \bra{\psi_2},
\end{equation}
where $p = {\cal N}^2 / (4 {\cal N}_1^2)$ and 
\begin{eqnarray}
\label{three-tangle-2}
&&\ket{\psi_1} = {\cal N}_1 \left[ (a - b) \ket {001} + 2 c \ket{010} + (a + b) \ket{111} \right]      \\    \nonumber
&&\ket{\psi_2} = {\cal N}_2 \left[ (a + b) \ket{000} + 2 \ket{011} + 2 c \ket{101} + (a -b) \ket{110}  \right].
\end{eqnarray}
The residual entanglements of $\ket{\psi_1}$ and $\ket{\psi_2}$ are 
\begin{equation}
\label{three-tangle-3}
\tau_3 (\psi_1) = 0    \hspace{1.0cm}
\tau_3 (\psi_2) = 64 {\cal N}_2^4 |(a^2 - b^2) c|.
\end{equation}

In order to compute the three-way entanglements of $\rho_{123}$ we consider the superposed state
\begin{equation}
\label{three-tangle-4}
\ket{\Psi (p, \varphi)} = \sqrt{p} \ket{\psi_1} + e^{i \varphi} \sqrt{1 - p} \ket{\psi_2}.
\end{equation}
If the phase factor $\varphi$ is chosen as 
\begin{equation}
\label{three-tangle-5}
\varphi = \varphi_{\pm} = - \frac{\theta_1 - \theta_2}{2} \pm \frac{\pi}{2}
\end{equation}
with $\theta_1 = \mbox{Arg} [(a^2 - b^2) c ]$ and $\theta_2 = \mbox{Arg} [(a^2 - c^2) (b^2 - c^2)]$, the residual entanglement 
of $\ket{\Psi (p, \varphi)}$ becomes
\begin{equation}
\label{three-tangle-6}
\tau_3 \left( \Psi (p, \varphi_{\pm}) \right) = 64 {\cal N}_1^2 {\cal N}_2^2 (1 - z) |(a^2 - c^2) (b^2 - c^2)| (1 - p) |p - p_0|
\end{equation}
where 
\begin{equation}
\label{three-tangle-7}
z = - \frac{{\cal N}_2^2}{{\cal N}_1^2} \left| \frac{(a^2 - b^2) c}{(a^2 - c^2) (b^2 - c^2)} \right|    \hspace{1.0cm}
p_0 = \frac{z}{z - 1}.
\end{equation}
Since $z \leq 0$, we get $ 0 \leq p_0 \leq 1$. Thus, $\tau_3 \left(\Psi (p, \varphi_{\pm}) \right)$ becomes zero at $p = p_0$.
It is easy to show that at the region $0 \leq p \leq p_0$ the sign of the second derivative of $\tau_3 \left( \Psi (p, \varphi_{\pm}) \right)$ becomes
\begin{equation}
\label{three-tangle-8}
\frac{d^2} {dp^2} \tau_3 \left( \Psi (p, \varphi_{\pm}) \right) \geq 0      \hspace{1.0cm}
\frac{d^2}{dp^2} \tau_3^{\frac{1}{q}} \left( \Psi (p, \varphi_{\pm}) \right) \leq 0    \hspace{.3cm} (q = 2, 3, 4, \cdots).
\end{equation}
Since the three-way entanglement  $t_{1|2|3}$ should be convex in the entire range of $p$, we have to adopt an appropriate convexification procedure 
appropriately. For, example, the optimal decomposition of $t^{(1)}_{1|2|3}$ is 
\begin{eqnarray}
\label{three-tangle-9}
\rho_{123} (p) = \left\{                                \begin{array}{c}
                                    \frac{p}{2 p_0} \left[ \ket{\Psi (p_0, \varphi_+ ) } \bra{\Psi (p_0, \varphi_+ ) }  + \ket{\Psi (p_0, \varphi_- ) } \bra{\Psi (p_0, \varphi_- ) }  \right] 
                                    + \frac{p_0 - p}{p_0} \ket{\psi_2} \bra{\psi_2}                                                                                 \\
                                \hspace{8.0cm}   (0 \leq p \leq p_0)                                                                                                                               \\
                                \frac{1 - p}{2 (1 - p_0)} \left[ \ket{\Psi (p_0, \varphi_+ ) } \bra{\Psi (p_0, \varphi_+ ) }  + \ket{\Psi (p_0, \varphi_- ) } \bra{\Psi (p_0, \varphi_- ) }  \right] 
                                + \frac{p - p_0}{1 - p_0} \ket{\psi_1} \bra{\psi_1}.                                                                                                         \\
                                 \hspace{8.0cm}   (p_0 \leq p \leq 1)
                                                                   \end{array}                                                                         \right.
\end{eqnarray}
The resulting $t^{(1)}_{1|2|3}$ is 
\begin{eqnarray}
\label{three-tangle-10}
 t^{(1)}_{1|2|3} = \left\{                          \begin{array}{cc}
                                            \left[ 8 {\cal N}_2^2 \sqrt{|(a^2 - b^2) c|} \left(1 - \frac{p}{p_0} \right)         \right]^{\mu_1}  &  \hspace{1.0cm}   (0 \leq p \leq p_0)         \\
                                            0  &  \hspace{1.0cm}   (p_0 \leq p \leq 1).
                                                                     \end{array}                           \right.
\end{eqnarray}

The optimal decomposition for $t^{(2)}_{1|2|3}$ at the region $0 \leq p \leq p_0$  is different from Eq. (\ref{three-tangle-9}) as 
\begin{equation}
\label{three-tangle-11}
\rho_{123} (p) = \frac{1}{2} \left[ \ket{\Psi (p, \varphi_+ ) } \bra{\Psi (p, \varphi_+ ) }  + \ket{\Psi (p, \varphi_- ) } \bra{\Psi (p, \varphi_- ) }  \right] 
\end{equation}
and the resulting $t^{(2)}_{1|2|3}$ becomes
\begin{eqnarray}
\label{three-tangle-12}
 t^{(2)}_{1|2|3} = \left\{                          \begin{array}{cc}
                                            \bigg( 64 {\cal N}_1^2 {\cal N}_2^2 (1 - z) |(a^2 - c^2) (b^2 - c^2)| (1 - p) (p_0 - p)       \bigg)^{\mu_2}  &  \hspace{1.0cm}   (0 \leq p \leq p_0)         \\
                                            0  &  \hspace{1.0cm}   (p_0 \leq p \leq 1).
                                                                     \end{array}                           \right.
\end{eqnarray}
The optimal decomposition for $t^{(3)}_{1|2|3}$ is exactly the same with that of $t^{(1)}_{1|2|3}$ and the resulting  $t^{(3)}_{1|2|3}$ is 
\begin{eqnarray}
\label{three-tangle-13}
 t^{(3)}_{1|2|3} = \left\{                          \begin{array}{cc}
                                             64 {\cal N}_2^4 |(a^2 - b^2) c| \left(1 - \frac{p}{p_0} \right)^q  &  \hspace{1.0cm}   (0 \leq p \leq p_0)         \\
                                            0  &  \hspace{1.0cm}   (p_0 \leq p \leq 1).
                                                                     \end{array}                           \right.
\end{eqnarray}
One can show straightforwardly that $t^{(a)}_{i|j|k} \hspace{.2cm} (a=1,2,3)$ of other three parties are the same with 
$t^{(a)}_{1|2|3}$ in the second class $\ket{G}$.

\section{Few special cases}

In this section we examine the minimal power which makes $\Delta_j$ to be positive when the power factors are larger than the corresponding minimal 
powers.

\subsection{special case I: $b = c = i a$}

%%%%%%%%%%%%%%%%%%%%%%%%%%%%%%%%%%%%%%%%%%%%%%%%%%%%%%%%%
\begin{figure}[ht!]
\begin{center}
\includegraphics[height=6.5cm]{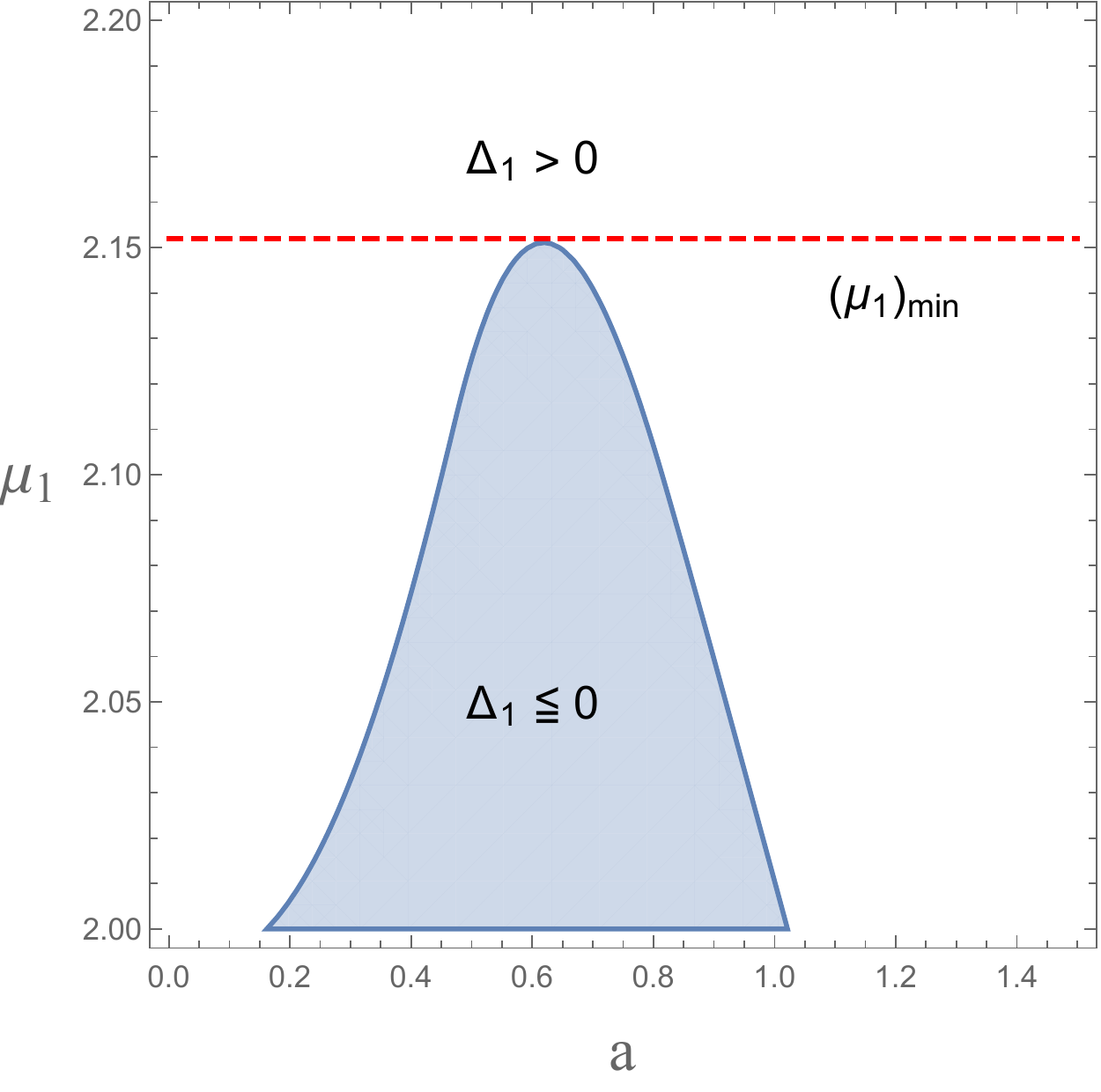}
\hspace{2.0cm}
\includegraphics[height=6.5cm]{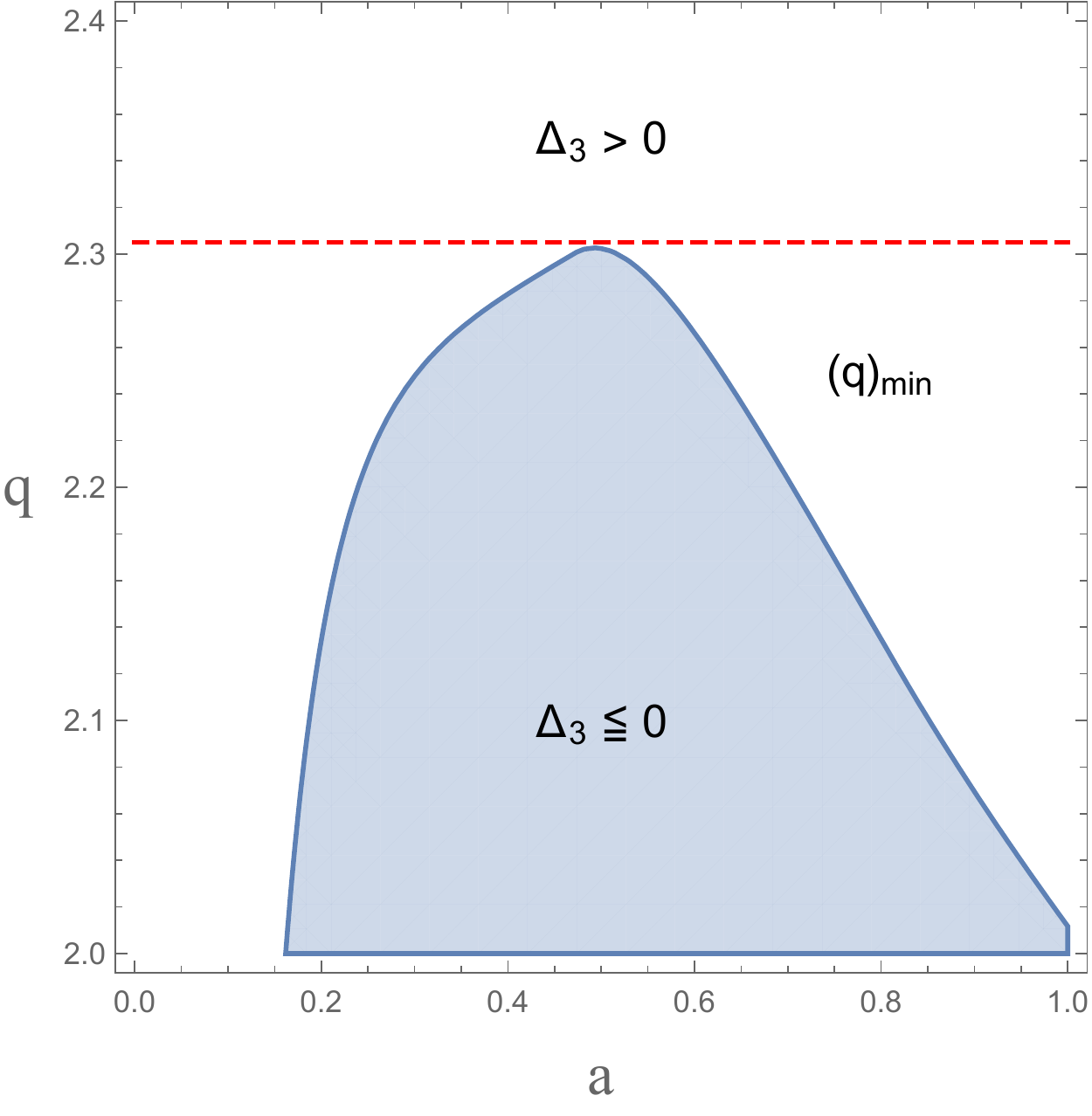}

\caption[fig1]{(Color online) The power factor dependence of the regions $\Delta_j > 0$ with varying $a$. 
 From Fig. 1(a)
 $\Delta_1$ and $\Delta_2$ become positive regardless of $a$ if $\mu_1 = 2 \mu_2 \geq (\mu_1)_{\min}$, where 
 $ (\mu_1)_{\min} = 2.152$. Fig. 1(b) shows that $\Delta_3$ becomes positive regardless of $a$ if $q \geq (q)_{\min}$, where $ (q)_{\min} = 2.305$.  }
\end{center}
\end{figure}
%%%%%%%%%%%%%%%%%%%%%%%%%%%%%%%%%%%%%%%%%%%%%%%%%%%%%%%%%%%

In this subsection we examine the minimal powers $(\mu_1)_{\min}$,  $(\mu_2)_{\min}$, $(q)_{\min}$, which make $\Delta_j$ positive when $a$ is 
positive and $b = c = i a$. In this case the normalization constants given in Eqs. (\ref{normalization1}) and (\ref{normalization2}) become
\begin{equation}
\label{special-1}
{\cal N} = \frac{1}{\sqrt{1 + 4 a^2}}     \hspace{.3cm}
{\cal N}_1 = \frac{1}{\sqrt{8} a}     \hspace{.3cm}    {\cal N}_2 = \frac{1}{2 \sqrt{1 + 2 a^2}}.
\end{equation}
Thus, the one-tangle $t_{1|234}$ simply reduces to 
\begin{equation}
\label{special-2}
t_{1|234} = \frac{8 a^2 (1 + 2 a^2)}{(1 + 4 a^2)^2}.
\end{equation}
Since Eq. (\ref{two-tangle-2}) yields
\begin{equation}
\label{special-3}
\alpha = - \alpha' = a^2       \hspace{.5cm}      \beta = \beta' = \frac{3}{2} a^2      \hspace{.5cm}
\gamma = \gamma' = \frac{a^2}{2}     \hspace{.5cm}    \delta^* = \delta' = \frac{1 + i}{2} a,
\end{equation}
the concurrences given in Eq. (\ref{two-tangle-3}) become
\begin{eqnarray}
\label{special-4}
&&{\cal C} (\rho_{12}) = \left\{          \begin{array}{cc}
                                       \frac{2 a (1 - a)}{1 + 4 a^2}   &   \hspace{2.0cm}   0 \leq a \leq 1                 \\
                                       0                                          &   \hspace{2.0cm}    1 \leq a
                                              \end{array}                                                                                           \right.               \\
&&{\cal C} (\rho_{13}) = {\cal C} (\rho_{14}) = \left\{             \begin{array}{cc}
                                                \frac{a}{1 + 4 a^2} (\sqrt{2} - 3 a)          &   \hspace{2.0cm}    0 \leq a \leq \frac{\sqrt{2}}{3}       \\
                             0                                                            &   \hspace{2.0cm}     \frac{\sqrt{2}}{3} \leq a \leq \sqrt{\frac{2}{3}}     \\
                             \frac{a}{1 + 4 a^2} (2 a - \sqrt{2 + a^2})  &    \hspace{2.0cm}      \sqrt{\frac{2}{3}} \leq a.
                                                                           \end{array}                                                               \right.
\end{eqnarray}
The parameters $p$, $z$, and $p_0$ defined in the previous section are given by $p = 2 a^2 / (1 + 4a^2)$, $z = -\infty$, and $p_0 = 1$ in this 
special case. Using these the various three-way entanglements become
\begin{equation}
\label{special-5}
t^{(1)}_{1|2|3} = \left( \frac{\sqrt{8 a^3}}{1 + 4 a^2} \right)^{\mu_1}      \hspace{1.0cm}
t^{(2)}_{1|2|3} = \left(  \frac{8 a^3}{(1 + 4 a^2)^2}  \right)^{\mu_2}          \hspace{1.0cm}
t^{(3)}_{1|2|3} = \frac{8 a^3 (1 + 2 a^2)^{q-2}}{(1 + 4 a^2)^q}.
\end{equation}
In this special case $t^{(1)}_{1|2|3} = t^{(2)}_{1|2|3}$ when $\mu_1 = 2 \mu_2$. However this relation does not hold generally.
Combining Eqs. (\ref{special-2}), ({\ref{special-4}), and (\ref{special-5}), one can compute $\Delta_j$ defined in Eq. (\ref{four-monogamy}), whose 
expressions are 
\begin{equation}
\label{special-6}
\Delta_j = -3 t^{(j)}_{1|2|3} + \frac{2 a^2} {1 + 4 a^2} f(a)
\end{equation}
where
\begin{eqnarray}
\label{special-7}
f(a) = \left\{                             \begin{array}{cc}
                                     a (4 + 6 \sqrt{2} - 3 a)     &    \hspace{2.0cm}    0 \leq a \leq \frac{\sqrt{2}}{3}          \\
                                     2 (1 + 2 a + 3 a^2)             &    \hspace{2.0cm}    \frac{\sqrt{2}}{3} \leq a  \leq \sqrt{\frac{2}{3}}       \\
                                     a (4 + a + 4 \sqrt{2 + a^2})    &    \hspace{2.0cm}        \sqrt{\frac{2}{3}} \leq a \leq 1               \\
                                     2 + 3 a^2 + 4 a \sqrt{2 + a^2}.     &    \hspace{2.0cm}    1 \leq a
                                               \end{array}                                                                                     \right.
\end{eqnarray}

 In Fig. 1 we plot the $\mu_1$-dependence of $\Delta_1 > 0$ region and $q$-dependence of $\Delta_3 > 0$ region with varying $a$. From Fig. 1(a)
 $\Delta_1$ and $\Delta_2$ become positive regardless of $a$ if $\mu_1 = 2 \mu_2 \geq (\mu_1)_{\min}$, where 
 $ (\mu_1)_{\min} = 2.152$. Fig. 1(b) shows that $\Delta_3$ becomes positive regardless of $a$ if $q \geq (q)_{\min}$, where $ (q)_{\min} = 2.305$. 

\subsection{special case II: $b = c > 0$}

%%%%%%%%%%%%%%%%%%%%%%%%%%%%%%%%%%%%%%%%%%%%%%%%%%%%%%%%%
\begin{figure}[ht!]
\begin{center}
\includegraphics[height=6.5cm]{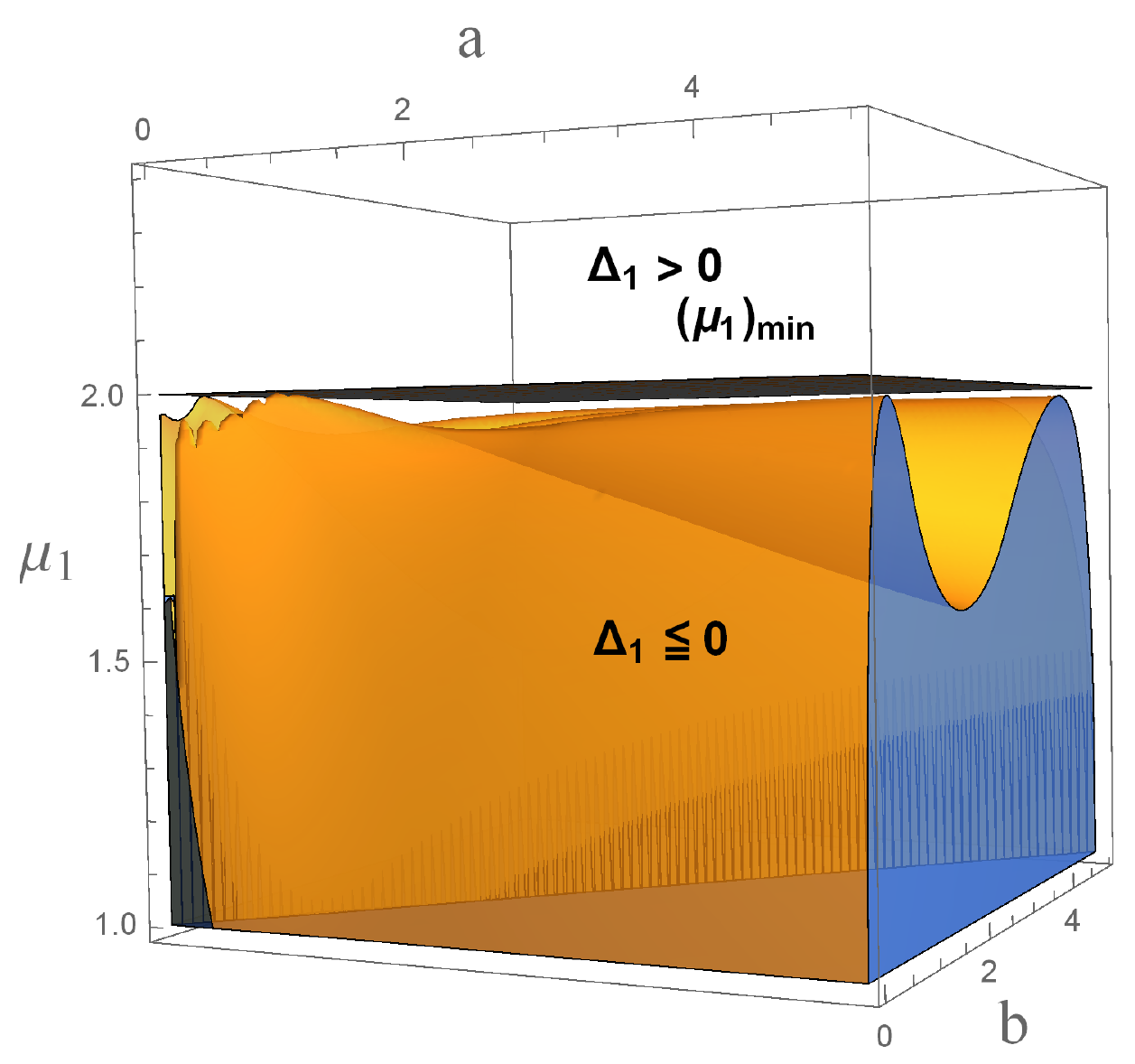}
\hspace{2.0cm}
\includegraphics[height=6.5cm]{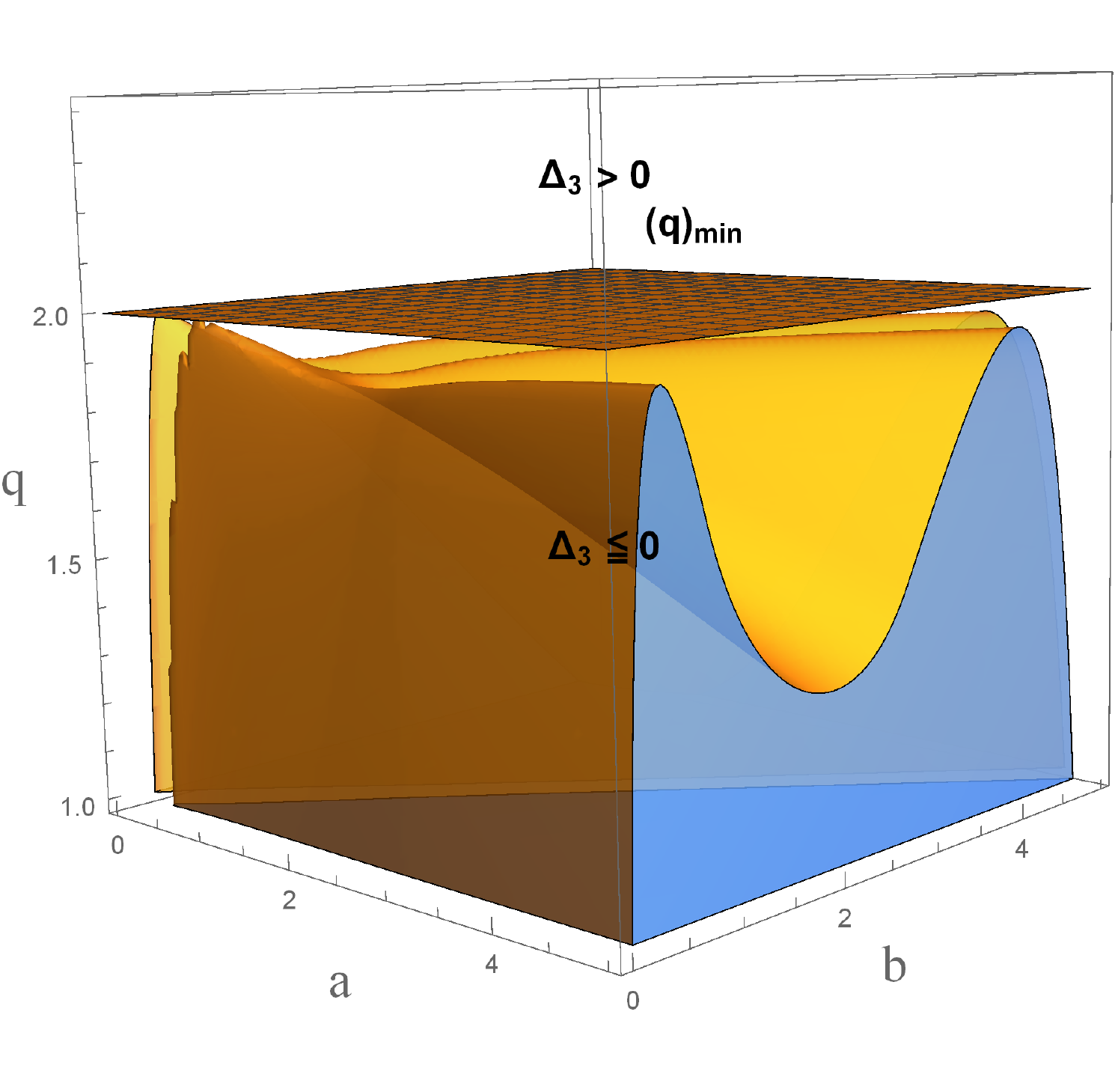}

\caption[fig2]{(Color online) The power factor dependence of the regions $\Delta_j > 0$ with varying $a$. 
 From Fig. 1(a)
 $\Delta_1$ and $\Delta_2$ become positive regardless of $a$ and $b$ if $\mu_1 = 2 \mu_2 \geq (\mu_1)_{\min}$, where 
 $ (\mu_1)_{\min} = 2.01$. Fig. 1(b) shows that $\Delta_3$ becomes positive regardless of $a$ and $b$ if $q \geq (q)_{\min}$, where $ (q)_{\min} = 2.00$.  }
\end{center}
\end{figure}
%%%%%%%%%%%%%%%%%%%%%%%%%%%%%%%%%%%%%%%%%%%%%%%%%%%%%%%%%%%

In this subsection we examine the minimal powers $(\mu_1)_{\min}$,  $(\mu_2)_{\min}$, $(q)_{\min}$, which make $\Delta_j$ positive when $a$, $b$,
and $c$ are real and positive with $b = c$. In this case the normalization constants given in Eqs. (\ref{normalization1}) and (\ref{normalization2}) become
\begin{equation}
\label{special-8}
{\cal N} = \frac{1}{\sqrt{1 +  a^2 + 3 b^2}}     \hspace{.3cm}
{\cal N}_1 = \frac{1}{\sqrt{2 (a^2 + 3 b^2)}}     \hspace{.3cm}    {\cal N}_2 = \frac{1}{ \sqrt{2 (2 + a^2 + 3 b^2)}}.
\end{equation}
Using the various formula presented in the previous section the one-tangle is given by
\begin{equation}
\label{special-9}
t_{1|234} = \frac{(a^2 + 3 b^2) (2 + a^2 + 3 b^2)}{(1 + a^2 + 3 b^2)^2}
\end{equation}
and the concurrences are 
\begin{eqnarray}
\label{special-10}
&&{\cal C} (\rho_{12}) = \left\{                                   \begin{array}{cc}
                                     {\cal N}^2 \max \left[ 2 b - (a^2 + b^2), 0 \right]  &  \hspace{2.0cm} b (\sqrt{1 + b^2} + 1)  \geq a^2     \\
                                     {\cal N}^2 \max \left[a^2 - b^2 - 2 b \sqrt{1 + b^2}, 0 \right]  &   \hspace{2.0cm} b (\sqrt{1 + b^2} + 1)  \leq a^2
                                                                       \end{array}                                    \right.                                   \\      \nonumber
&&{\cal C} (\rho_{13}) = \left\{                                   \begin{array}{cc}
                      2 {\cal N}^2 \max \left[ b (a + b) - \frac{|a - b|}{2} \sqrt{1 + \left( \frac{a - b}{2} \right)^2}, 0 \right]   &
                         \hspace{.2cm} \frac{(a + 3 b)^2}{4} \geq \frac{|a - b|}{2} \left( 1 + \sqrt{1 + \left( \frac{a - b}{2} \right)^2} \right)      \\
                     \frac{{\cal N}^2}{2} \max \left[ 2 |a - b| - (a^2 + 2 a b + 5 b^2), 0 \right]  &
                      \hspace{.2cm} \frac{(a + 3 b)^2}{4} \leq \frac{|a - b|}{2} \left( 1 + \sqrt{1 + \left( \frac{a - b}{2} \right)^2} \right)   
                                                                            \end{array}                                \right.                                   \\      \nonumber
&& 
{\cal C} (\rho_{14}) = \left\{                                   \begin{array}{cc}
 2 {\cal N}^2 \max \left[ b (b - a) - \frac{a + b}{2} \sqrt{1 + \left( \frac{a + b}{2} \right)^2}, 0 \right]   &              
                         b \geq a \hspace{.2cm}\mbox{and} \hspace{.2cm} \frac{(a - 3 b)^2}{4} \geq \frac{a + b}{2} \left( 1 + \sqrt{1 + \left( \frac{a + b}{2} \right)^2} \right)      \\
                     \frac{{\cal N}^2}{2} \max \left[ 2 (a + b) - (a^2 - 2 a b + 5 b^2), 0 \right]. &
                      \hspace{.2cm} \mbox{elsewhere}
                                                                            \end{array}                                \right. 
\end{eqnarray}
Since $p_0 = 1$ in this case too, the $t_{1|2|3}^{(j)}$ given in Eqs. (\ref{three-tangle-10}), (\ref{three-tangle-12}), and (\ref{three-tangle-13}) are 
expressed as 
\begin{eqnarray}
\label{special-11}
&&t_{1|2|3}^{(1)} = \left( \frac{2 \sqrt{b |a^2 - b^2|}}{1 + a^2 + 3 b^2}  \right)^{\mu_1}    \hspace{1.0cm}
t_{1|2|3}^{(2)} = \left( \frac{4 b |a^2 - b^2|}{(1 + a^2 + 3 b^2)^2} \right)^{\mu_2}                             \nonumber   \\
&&\hspace{2.0cm} t_{1|2|3}^{(3)} =2^{4 - q} b |a^2 - b^2| \frac{(2 + a^2 + 3 b^2)^{q - 2}}{(1 + a^2 + 3 b^2)^q}.
\end{eqnarray}
As in the previous special case we have a relation $t_{1|2|3}^{(1)} = t_{1|2|3}^{(2)}$ if $\mu_1 = 2 \mu_2$. 

In Fig. 2 the full parameter space is divided into two regions, i.e. $\Delta_j > 0$ and $\Delta_j \leq 0$ regions. The division enables us to find the 
minimal powers $(\mu_1)_{\min}$,  $(\mu_2)_{\min}$, and $q_{\min}$, which makes $\Delta_j \geq 0 \hspace{.2cm} (j=1,2,3)$ regardless of the 
parameters. Fig. 2 shows $(\mu_1)_{\min} = 2 (\mu_2)_{\min} = 2.01$ and $(q)_{\min} = 2.00$ in this special case.

\section{Numerical Analysis}
In this section we compute the minimal powers  $(\mu_1)_{\min}$,  $(\mu_2)_{\min}$, and  $(q)_{\min}$ for some more general cases by making use of 
numerical approach. First, we consider $b = c = i r a$ with $a > 0$. Since $p_0 = 1$ in this case, $t_{1|2|3}^{(j)}$ can be computed directly. One can show 
easily $t_{1|1|3}^{(1)} = t_{1|1|3}^{(2)}$ if $\mu_1 = 2 \mu_2$ in this case too. Thus, we have a constraint  $(\mu_1)_{\min} = 2 (\mu_2)_{\min}$. 

\begin{center}
\begin{tabular}{c|ccccccccccc} \hline \hline
 $r$ &  $0.1$  &  $0.2$  &  $0.3$  &  $0.4$  &  $0.5$  &  $0.6$  &  $0.7$  &  $0.8$  &  $0.9$  &  $1.0$  &  $10$     \\  \hline 
 $(\mu_1)_{\min}$ &  $1.99$  &  $2.01$  &  $2.07$  &  $2.17$  &  $2.26$  &  $2.27$  &  $2.25$  &  $2.21$  &  $2.18$  &  $2.15$  &  $2.00$     \\
 $(q)_{\min}$  & $1.97$  &  $2.01$  &  $2.09$  &  $2.31$  &  $3.31$  &  $13.6$  &  $2.91$  &  $2.41$  &  $2.33$  &  $2.31$  &  $2.00$      \\  \hline   \hline
\end{tabular}

\vspace{0.1cm}
Table I: The $r$-dependence of the minimal powers $(\mu_1)_{\min}$ and $(q)_{\min}$ when $b = c = i r a$.
\end{center}
The $r$-dependence of the minimal powers  $(\mu_1)_{\min}$ and $(q)_{\min}$ is summarized in Table I. Both powers increase with increasing $r$ 
from $0.1$ to $0.6$. Both decrease with increasing $r$ from $0.6$ and seem to be saturated to $(\mu_1)_{\min} = (q)_{\min} = 2$ at large $r$. At $r = 0.6$ 
$(q)_{\min} $ becomes very large as $(q)_{\min} = 13.6$ while  $(\mu_1)_{\min}$ is not so large as  $(\mu_1)_{\min} = 2.27$.

Next, we consider $b = i r a$ and $c = n r a$, where $r$ and $a$ are real with integer $n$. The minimal powers can be computed by making use of the 
three-dimensional plot similar to Fig. 2. The results are summarized in Table II.

\begin{center}
\begin{tabular}{c|ccccc} \hline \hline
 $n$ &  $1$  &  $2$  &  $3$  &  $4$  &  $5$   \\  \hline 
 $(\mu_1)_{\min}$ &  $2.13$  &  $2.03$  &  $2.01$  &  $2.00$  &  $2.00$      \\
 $(\mu_2)_{\min}$  &  $1.07$  &  $1.02$  &  $1.003$  &  $1.00$  &  $0.99$     \\
 $(q)_{\min}$  & $2.28$  &  $2.10$  &  $1.98$  &  $1.85$  &  $1.91$      \\  \hline   \hline
\end{tabular}

\vspace{0.1cm}
Table II: The $n$- dependence of minimal powers  when  $b = i r a$ and $c = n r a$.
\end{center}
All minimal powers exhibit decreasing behavior with increasing $n$. In this case $(\mu_2)_{\min}$ roughly equals to the half of  $(\mu_1)_{\min}$ as
in the previous cases.

We also examine the case of $b = n i$ when $a$ and $c$ are real. The minimal powers of this case is summarized in Table III.

\begin{center}
\begin{tabular}{c|ccccc} \hline \hline
 $n$ &  $1$  &  $2$  &  $3$  &  $4$  &  $5$   \\  \hline 
 $(\mu_1)_{\min}$ &  $2.28$  &  $2.03$  &  $1.99$  &  $1.99$  &  $1.99$      \\
 $(\mu_2)_{\min}$  &  $1.14$  &  $1.02$  &  $0.98$  &  $0.98$  &  $0.98$     \\
 $(q)_{\min}$  & $2.36$  &  $2.04$  &  $1.98$  &  $1.97$  &  $1.97$      \\  \hline   \hline
\end{tabular}

\vspace{0.1cm}
Table III: The $n$- dependence of minimal powers  when  $b = n i$. 
\end{center}
Similar to the previous case all minimal powers exhibit decreasing behavior with increasing $n$. In this case also $(\mu_2)_{\min}$ roughly equals to the half of  $(\mu_1)_{\min}$.

Finally, we choose $N = 10000000$ second class states randomly with imposing $b=c$ and compute $\Delta_j$ with particular powers. The number of states
which give negative $\Delta_1$ or $\Delta_2$ are summarized in Table IV. 

\begin{center}
\begin{tabular}{c||cccc|cccc} \hline \hline
 $\mu_1$ or $\mu_2$ &  \hspace{.2cm} $\mu_1 =2.0$  &  $2.1$  &  $2.2$  &  $2.3$ \hspace{.2cm} & \hspace{.2cm} $\mu_2 = 1.00$  &  $1.05$  &  $1.10$  &  $1.15$    \\    \hline
 No. states  & $1216071$  &  $191610$  &  $16818$  &  $0$  &  $1213371$  &  $191002$  &  $16755$  &  $0$     \\  \hline   \hline
\end{tabular}

\vspace{0.1cm}
Table IV:  Number of states which give negative $\Delta_1$ or $\Delta_2$ for arbitrary chosen 10000000 states.
\end{center}
The number of states which give negative $\Delta_3$ are summarized in Table V.

\begin{center}
\begin{tabular}{c|cccccccc} \hline \hline
 $q$ &   $2.0$  &  $2.1$  &  $2.2$  &  $2.3$  &  $2.4$  &  $2.5$  &  $2.6$  &  $2.7$    \\    \hline
 No. states  & $1214527$  &  $429823$  &  $170308$  &  $49247$  &  $832$  &  $110$  &  $35$  &  $7$     \\  \hline   \hline
\end{tabular}

\vspace{0.1cm}
Table V:  Number of states which give negative $\Delta_3$  for  arbitrary chosen 10000000 states.
\end{center}

All the results discussed in section II and section III indicate that $(\mu_1)_{\min} \approx 2 (\mu_2)_{\min}  \geq 2.3$ and 
$(q)_{\min} \geq 14$, at least in the whole second class. However, as Table I and Table V indicate, the region of negative $\Delta_3$ in the parameter space 
is extremely small for $2.7 \leq q \leq 13$. Thus, it seems to be highly difficult to find such states in the random number generation.

\section{Four-Way Entanglement Measure}
In this section we discuss a following question: Is it possible that the monogamy relation $\Delta_j (G)$ defined in Eq. (\ref{four-monogamy}) quantifies
the SLOCC-invariant  four-way entanglement in particular powers like a leftover of CKW inequality in three-way entanglement? 
In order to explore this question we note that for $n$-qubit system there are 
$2(2^n - 1) - 6 n$ independent SLOCC-invariant monotones\cite{verst03}. 
Thus, in four-qubit system there are six invariant monotones. Among them, it was shown in 
Ref. \cite{four-way-1,four-way-2,four-way-3} by making use of the antilinearity\cite{uhlmann99-1} that there are following three independent invariant monotones which measure the true four-way entanglement:
\begin{eqnarray}
\label{four-measure}
& &{\cal F}^{(4)}_1 = (\sigma_{\mu} \sigma_{\nu} \sigma_2 \sigma_2) \bullet (\sigma^{\mu} \sigma_2 \sigma_{\lambda} \sigma_2) \bullet
                      (\sigma_2 \sigma^{\nu} \sigma^{\lambda} \sigma_2)           \nonumber  \\
& &{\cal F}^{(4)}_2 = (\sigma_{\mu} \sigma_{\nu} \sigma_2 \sigma_2) \bullet (\sigma^{\mu} \sigma_2 \sigma_{\lambda} \sigma_2) \bullet (\sigma_2 \sigma^{\nu} \sigma_2 \sigma_{\tau}) \bullet (\sigma_2 \sigma_2 \sigma^{\lambda} \sigma^{\tau})                                                                 \\    \nonumber
& &{\cal F}^{(4)}_3 = \frac{1}{2} (\sigma_{\mu} \sigma_{\nu} \sigma_2 \sigma_2) \bullet (\sigma^{\mu} \sigma^{\nu} \sigma_2 \sigma_2) \bullet (\sigma_{\rho} \sigma_2 \sigma_{\tau} \sigma_2) \bullet
(\sigma^{\rho} \sigma_2 \sigma^{\tau} \sigma_2) \bullet (\sigma_{\kappa} \sigma_2 \sigma_2 \sigma_{\lambda})
\bullet (\sigma^{\kappa} \sigma_2 \sigma_2 \sigma^{\lambda}),
\end{eqnarray}
where $\sigma_0 = \openone_2$, $\sigma_1 = \sigma_x$, $\sigma_2 = \sigma_y$, $\sigma_3 = \sigma_z$, and
the Einstein convention is understood with a metric $g^{\mu \nu} = \mbox{diag} \{-1, 1, 0, 1\}$. 
The solid dot in Eq. (\ref{four-measure}) is defined as follows. Let $\ket{\psi}$ be a four-qubit state. Then, for example, ${\cal F}^{(4)}_1$ of $\ket{\psi}$ is defined as 
\begin{equation}
\label{revise1}
{\cal F}^{(4)}_1 (\psi) = \bigg| \bra{\psi^*} \sigma_{\mu} \otimes \sigma_{\nu} \otimes \sigma_2 \otimes \sigma_2 \ket{\psi} \bra{\psi^*} \sigma^{\mu} \otimes \sigma_2 \otimes \sigma_{\lambda} \otimes \sigma_2 \ket{\psi}
                                                  \bra{\psi^*} \sigma_2 \otimes \sigma^{\nu} \otimes \sigma^{\lambda} \otimes \sigma_2  \ket{\psi}   \bigg|.
\end{equation}
Of course, other measures can be computed similarly. Thus, if $\Delta_j (G)$ properly quantifies the SLOCC-invariant four-way entanglement, 
it should be represented as a 
combination of ${\cal F}_j^{(4)}$. For simplicity, we consider only the second class state (\ref{second-class}) with $b = c = i a$. In this case $\Delta_j$ 
is computed analytically in Eq. (\ref{special-6}). In this case ${\cal F}^{(4)}_j$ becomes
\begin{equation}
\label{four_measure-1}
{\cal F}^{(4)}_1 = \frac{48 a^6}{(1 + 4 a^2)^3}     \hspace{.5cm}
{\cal F}^{(4)}_2 = \frac{96 a^8}{(1 + 4 a^2)^4}     \hspace{.5cm}
{\cal F}^{(4)}_3 = \frac{3456 a^{12}}{(1 + 4 a^2)^6}.
\end{equation}   
%%%%%%%%%%%%%%%%%%%%%%%%%%%%%%%%%%%%%%%%%%%%%%%%%%%%%%%%%
\begin{figure}[ht!]
\begin{center}
\includegraphics[height=8.0cm]{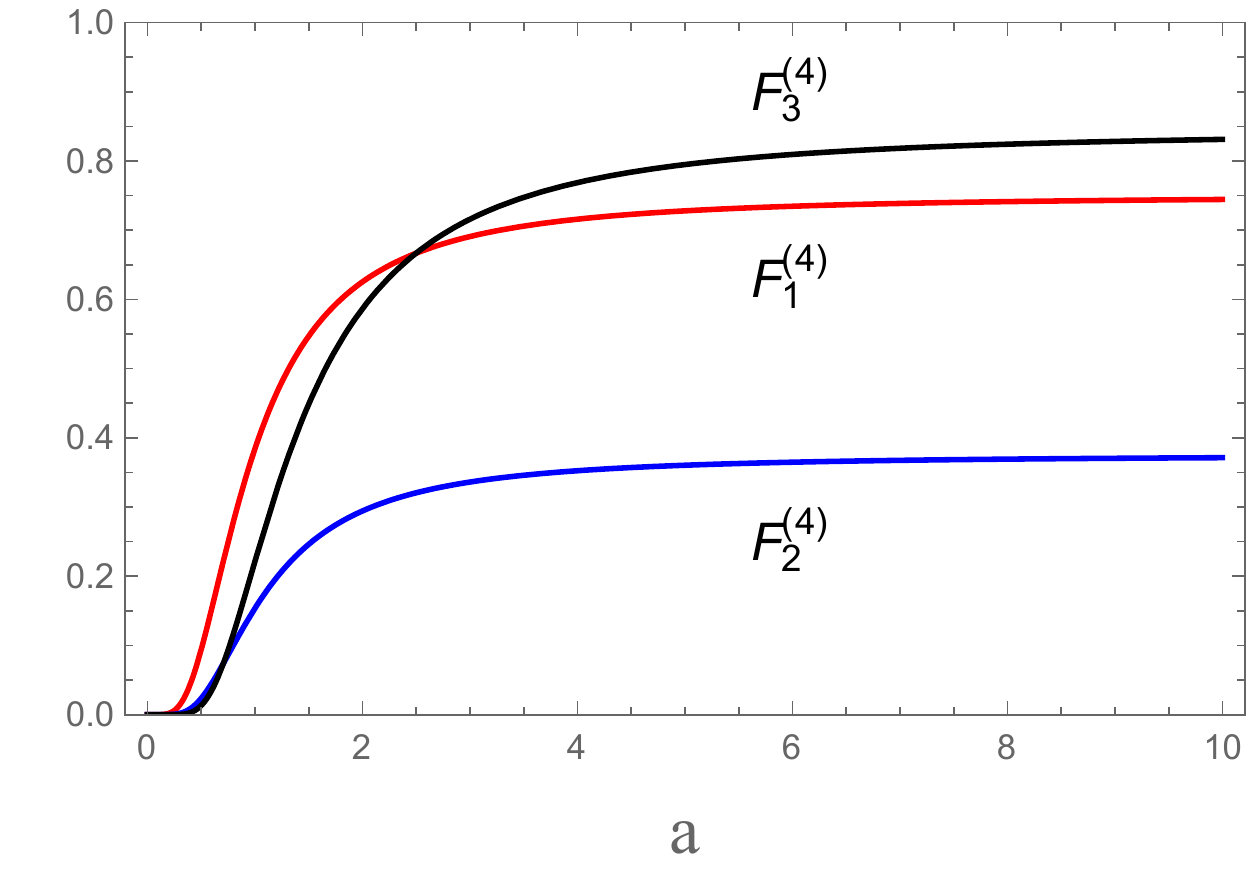}
%\hspace{2.0cm}
%\includegraphics[height=6.5cm]{fig2b.pdf}

\caption[fig3]{(Color online) The $a$-dependence of the four-way entanglement measures ${\cal F}^{(4)}_j$ for the second class state (\ref{second-class}) with imposing $b = c = i a$.  }
\end{center}
\end{figure}
%%%%%%%%%%%%%%%%%%%%%%%%%%%%%%%%%%%%%%%%%%%%%%%%%%%%%%%%%%%
These results are plotted in Fig. 3. 
Thus, all four-way entanglement measures ${\cal F}^{(4)}_j$ exhibit monotonically increasing behavior with respect to $a$ when the quantum state is chosen 
as a second class (\ref{second-class}) with $b = c = i a$. 

It is easy to show that $\Delta_j$ cannot be expressed in terms of  ${\cal F}^{(4)}_j$ because $\Delta_j$ have different expressions in the various 
 range of $a$ as Eq. (\ref{special-7}) shows while ${\cal F}^{(4)}_j$ have same expressions regardless of the range of $a$. For example, one can find 
a least-square fit of $\Delta_1$ with $\mu_1 = 3$ as 
\begin{equation}
\label{fit1}
\Delta_1 (\mu_1 = 3) \approx c_1  {\cal F}^{(4)}_1 + c_2 {\cal F}^{(4)}_2 + c_3 {\cal F}^{(4)}_3
\end{equation}
where $c_1 = 10.117$, $c_2 = -30.8143$, and $c_3 = 5.7116$.
%%%%%%%%%%%%%%%%%%%%%%%%%%%%%%%%%%%%%%%%%%%%%%%%%%%%%%%%%
\begin{figure}[ht!]
\begin{center}
\includegraphics[height=8.0cm]{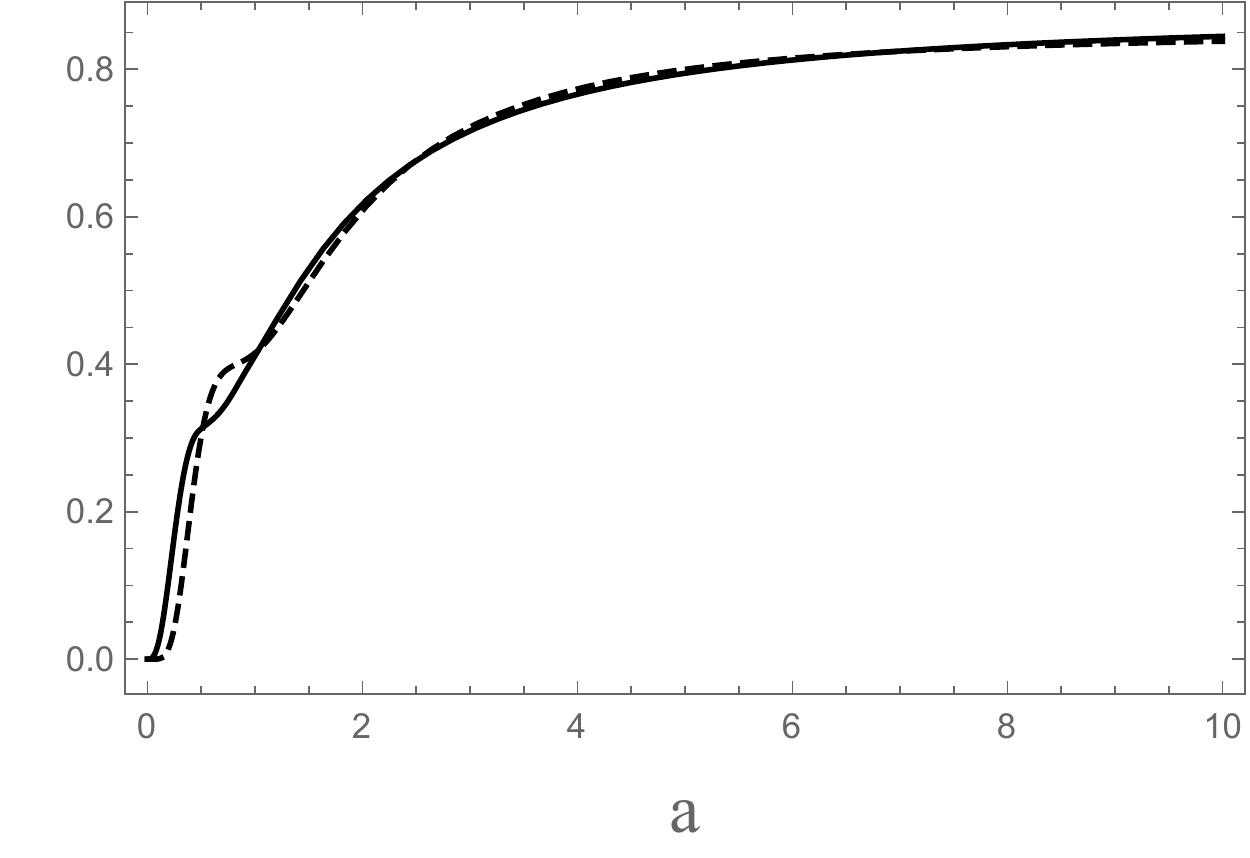}
%\hspace{2.0cm}
%\includegraphics[height=6.5cm]{fig2b.pdf}
\caption[fig4]{(Color online) The $a$-dependence of the  left- and right-handed sides of Eq. (\ref{fit1}). The former and latter are plotted as solid and dashed
lines respectively. This figure indicates that the monogamy constraint $\Delta_1$ cannot properly quantify the SLOCC-invariant  four-way entanglement.  }
\end{center}
\end{figure}
%%%%%%%%%%%%%%%%%%%%%%%%%%%%%%%%%%%%%%%%%%%%%%%%%%%%%%%%%%%
The left- and right-handed sides of Eq. (\ref{fit1}) are plotted in Fig. 4 as solid and dashed lines. Although both exhibit similar behavior, they do not 
coincide with each other exactly as expected. Same is true for $\Delta_2$ and $\Delta_3$. Thus, the monogamy constraints (\ref{four-monogamy-1})
derived by introducing a weighting factor in the power of the three-way entanglement cannot quantify the four-way entanglement properly in the SLOCC-invariant 
manner.

\section{Conclusions}
In this paper we examine the properties of the three four-qubit monogamy relations presented in Eq. (\ref{four-monogamy}), all of which introduce the 
power factors $\mu_1$, $\mu_2$, and $q$ in the three-way entanglement. First, we examine the minimal powers $(\mu_1)_{\min}$, 
 $(\mu_2)_{\min}$and  $(q)_{\min}$, which make $\Delta_j  \hspace{.3cm} (j=1, 2, 3)$ to be positive when the powers are larger than the minimal 
 powers. In order to explore this problem on the analytic ground as much as possible we confine ourselves into the second-class state $\ket{G}$ 
 defined in Eq. (\ref{second-class}). Our analysis indicates that $(\mu_1)_{\min} \approx 2 (\mu_2)_{\min} \geq 2.3$ and $(q)_{\min} \geq 14$.
 
 Second, we try to provide an answer to a question ``can the leftovers of the four-qubit monogamy relations with particular powers be a SLOCC-invariant 
 four-way entanglement measures like that of CKW inequality in three-qubit system?''. Our analysis indicates that this is impossible in the monogamy relations
 given in Eq. (\ref{four-monogamy}). Probably, same is true if monogamy relation is derived by introducing any form of  weighting factors. 
 Then, it is natural to ask a following question: Does the monogamy inequality exist in the multipartite qubit system, whose leftover quantifies the 
 SLOCC-invariant entanglement measure? We do not have definite answer to this question.

{\bf Acknowledgement}:
%On April 16, 2014 the ferry Sewol has sunk into the South Sea of Korea. Due to this disaster 304 people died and, 9 of them are still missing. We would like to dedicate this paper to all victims of this accident.
%This research was supported by the Basic Science Research Program through the National Research Foundation of Korea(NRF) funded by the Ministry of Education, Science and Technology(2011-0011971).
This work was supported by the Kyungnam University Foundation Grant, 2016.

\end{document}